\documentclass[
    ,final            % use final for the camera ready runs
  ]
  {aipproc}

\layoutstyle{6x9}

\begin{document}

\title{The jigsaw puzzle of scalar mesons}


\author{M. Boglione}{
  address={Dipartimento di Fisica Teorica, Universita' di Torino, 
           via P. Giuria 1, 10125 Italy }}

\begin{abstract}
This is a  brief overview of light scalar meson spectroscopy, addressing   
longstanding problems, recent developments and future perspectives.
In particular, a new comprehensive data analysis is introduced which will 
help to unravel the structure of the $f_0(980)$.
\end{abstract}

\maketitle

We all know that the quark model works well for most mesons: nice nonet 
structures arise when all possible combinations of $q\overline q$ pairs are 
ordered according to their isospin and strangeness.
Then, by exploiting the mass and decay properties of the 
physical mesons delivered by experiments, we can find a slot for each 
candidate in the nonet.
This game can be safely played for vector and tensor mesons and, to some 
extent, for pseudoscalar mesons, if the appropriate mixing angles are taken 
into account.
Consider, for instance, the $\omega(782)$ and the $\phi(1020)$: experiments 
tell us that the $\omega(782)$ decays mostly into pions and 
is lighter than the  $\phi(1020)$ which, on the contrary, decays into 
$K\overline K$ $85$\% of the time. It's mass being close to that of the 
$\rho(770)$ provides clear indication that the $\omega(782)$ is the $I=0$ 
non-strange candidate, whereas  $\phi(1020)$ is undoubtedly the $I=0$ 
$s\overline s$ member of the vector nonet. 
Similarly for the tensors $f_2$ and $f_2^\prime$.

For scalar mesons this does not work. The quark model fails inexorably: first 
of all, experiments detect many more physical scalar resonances than can fit 
in a nonet. Secondly, their decay properties are mostly unknown, so there is 
little guide to their classification, thirdly their spectra cannot be 
approximated by Breit-Wigner shapes, because they overlap and interfere with 
each other, some of them being very broad.
Therefore, the classical methods of analysing data cannot be applied.

How can we try and disentagle such a complicated picture?
{\it Unitarity} comes to our rescue. Indeed, this 
property, which follows from conservation of probability, has  
to be fulfilled whatever the quantum numbers of the $q\overline q$ pair, and 
give very useful constraints for our analyses. 
Unitarity requires the $T$ matrix for each partial wave to satisfy 
\({\rm Im} T = \rho |T|^2\), where $\rho$ is the phase space matrix. 
This relation constrains the imaginary part of $(1/T)$ to be  
\({\rm Im} (1/T) = -\rho\), in the simplest case, leaving  \({\rm Re} (1/T)\) 
unconstrained. By parametrizing \({\rm Re} (1/T)\) by a real 
matrix $1/K$, one obtains \(T=\frac{K}{1-i\rho K}\), which is the usual 
$K$-matrix representation. 
If there is only one channel, like in $\pi\pi\to\pi\pi$ scattering below 
$K\overline K$ threshold, and only one narrow resonance, this resonance will 
appear like a single pole in the $K$ amplitude, \(K=\frac{g^2}{M^2-s}\), and 
the T amplitude can be approximated by \(T=\frac{g^2}{M^2-s-i\rho g}\). 
The pole of $K$ gives the ``bare state'' and $T$ has a Breit-Wigner form.
This simple picture works only for narrow and well separated resonances, 
where coupling to hadronic loops has little effect. For 
the scalar sector, where resonances are broad (i.e. their poles are located 
very far from the real $s$-axis, where experiments happen), interfeering and 
overlapping (i.e. their spectra are not made of nicely separated peaks), 
this simple interpretation breaks down. 

\begin{figure}[t]
\includegraphics[height=.32\textheight]{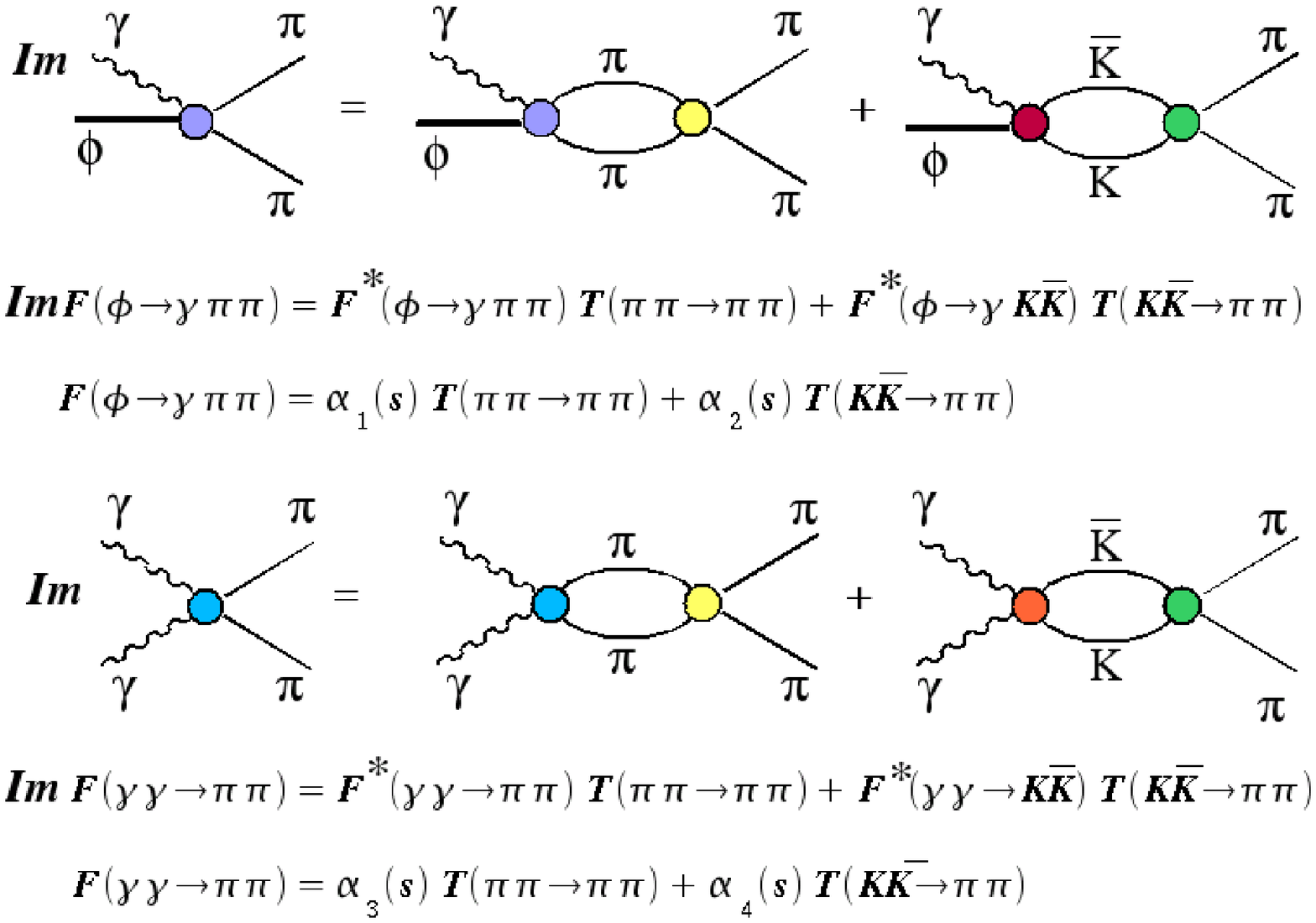}
\caption{\label{unitarity} Coupled channel unitarity constrains the amplitudes 
$F(\phi\to\gamma\pi\pi)$ and $F(\gamma\gamma\to\pi\pi)$ in terms of hadronic 
amplitudes corresponding to final state interactions, $\pi\pi\to\pi\pi$ and 
$K\overline K\to\pi\pi$ for a given $I$, $J$. 
For $\phi$-decay, the photon is assumed to be a spectator.}
\end{figure}

Fig.~\ref{unitarity} shows how similarly coupled-channel unitarity 
constrains the partial wave amplitudes $F$ corresponding to two different 
processes $\phi\to\gamma\pi\pi$ and $\gamma\gamma\to\pi\pi$; 
scalar meson resonances are produced in the final state 
interactions $\pi\pi\to\pi\pi$ and $K\overline K\to\pi\pi$ and are embodied 
as poles in the $I=J=0$ hadronic amplitudes, $T$.  
The general solution of the unitarity requirement for the $F$'s is given by a 
linear combination of the $T$'s, where the coefficients 
$\alpha _i(s)$ are real functions of $s$, simple polynomials  
apart from some factors as explained in \cite{BP-phi}. 
Notice that unitarity requires {\it consistency} between reactions, in that 
the same strong interaction amplitudes $T$, combined and weighted using 
appropriate $\alpha _i$ coefficients, form the amplitudes 
corresponding to different reactions. The $\alpha$-vector formulation embodies 
{\it universality}, demanding that poles of the $S$ matrix transmit to all 
processes with the same quantum numbers in exactly the same position. 
This indeed makes the determination of the $F$ amplitudes very sensitive to 
the details of the $T$'s.
%%%%%%%%%%%%%%%%%%%%%%%%%%%%%%%%%%%%%%%%%%%%
%%%%%%%%%%%%%%%%%%%%%%%%%%%%%%%%%%%%%%%%%%%%

Recently, M.R. Pennington and I made an analysis \cite{BP-phi} of 
$\phi\to\gamma\pi\pi$ experimental data \cite{kloe} based on the 
coupled channel unitarity constrains of Fig.~\ref{unitarity} and showed that 
huge differences arise in the determination of the relevant couplings and the 
$\phi\to\gamma f_0(980)$ branching ratio due to different choices of 
underlying amplitudes $T$. We chose an old set of 
hadronic amplitudes called ReVAMP, determined as in \cite{MP} and a 
recent one, obtained by Anisovich and Sarantsev in \cite{AS} fitting a much 
larger amount of data. In the first set of amplitudes, the $f_0(980)$ appears 
as a narrow resonance, lighter than the $\phi(1020)$. 
In the second case the $f_0(980)$ is a much broader object, heavier than the 
$\phi(1020)$. Since the decay rate 
distribution depends crucially on the {\it cube} of the photon momentum, i.e.  
$(m_\phi^2-s)^3$, and since the $f_0(980)$ is so close to the end of phase 
space, it turns out that the determination of the couplings and 
branching ratio is extremely sensitive to the exact position of the 
$f_0(980)$ pole in the $T$'s. 
The fit clearly favours the ReVAMP set of amplitudes, which give an excellent 
quality of results with constant $\alpha _i(s)$ (3 parameter fit), confirming 
that the $\pi\pi$ final state interactions in this particular process are 
consistent with those of the processes exploited to determine the ReVAMP 
amplitudes. Indeed, when the new, high statistics, KLOE data will be released, 
we will have the chance to test this consistency further.

While for decays like $\phi\to\pi\pi X$ we have to assume $X$ is a spectator 
to apply unitarity as in Fig.1, for  $\gamma\gamma\to\pi\pi$ scattering 
unitarity and universality apply with no assumptions. 
A few years ago, M.R. Pennington and I analysed $\gamma\gamma\to\pi\pi$ 
world data \cite{BP-gamma} to determine the radiative widths of scalar mesons. 
The underlying hadronic amplitudes we used were the same ReVAMP set described 
above.
We found two classes of solutions, delivered by fits equally good in quality 
and giving comparable scalar widths: one where the $f_0(980)$ showed up as a 
peak, and the other where the $f_0(980)$ showed up as a dip. 
Shortly, new very high statistics data from BELLE and BaBar will be available:
they will allow us a global reanalysis to discern between the two solutions 
and to test the $T$ underlying hadronic amplitudes. 

For these re-analysis, we are considering a different parametrization 
for the $T$'s.  
In fact, the simplest solution to the unitarity requirement, as shown above,
 violates left hand cut analyticity: each $\rho$ matrix element is singular at 
$s\to \infty$, which constrains the $T$'s in an artificial and unnecessary way.
To avoid this, we perform new fits \cite{BP-FSI} that include recent 
experimental data in addition to those used for the original ReVAMP analysis, 
in which ${\rm Im} (1/T)$ is given by the Chew-Mandelstam function, which is 
not affected by that flaw.     

Concluding, the main message of this talk is the following: unitarity and 
analyticity give powerful constraints and must be at the very basis of any 
data analysis. 
Unitarity requires {\it consistency} among different reactions, so that 
analysing data where final state interactions are important only makes sense 
if it is done in a global and comprehensive way. It's like a big jigsaw 
puzzle game: you have to take care of combining appropriately all the single 
pieces before the total picture is revealed.

\begin{theacknowledgments}
It is a pleasure to thank Umberto D'Alesio for inviting me to this conference,
and to all the organizers for their warmest hospitality. 
I am infinitely grateful to M.R. Pennington for many invaluable  
discussions on this subject and for years of fruitful and enjoyable 
collaboration.
\end{theacknowledgments}
%



\bibliographystyle{aipproc}   % if natbib is available

%\bibliography{sample}

%  \typeout{******************************************}
%  \typeout{** Please run "bibtex \jobname" to obtain}
%  \typeout{** the bibliography and then re-run LaTeX}
%  \typeout{** twice to fix the references!}
%  \typeout{******************************************}

%\end{document}

%%%%%%%%%%%%%%%%%%%%%%%%%%%%%%%%%%%%%%%%%%%
%% The following lines show an example how to produce a bibliography
%% without the help of the BibTeX program. This could be used instead
%% of the above.
%%%%%%%%%%%%%%%%%%%%%%%%%%%%%%%%%%%%%%%%%%%

\end{document}